\documentclass[prl,aps,showpacs,twocolumn,floatfix]{revtex4-1}
\usepackage[all]{xy}
\usepackage{amsmath,amsfonts,amssymb,graphicx}

\newcommand{\uf}{U_{\!f}}

\def\openone{\leavevmode\hbox{\small1\kern-3.8pt\normalsize1}}
\def\RR{{\rm I\kern-.2emR}}

\def\openone{\leavevmode\hbox{\small1\kern-3.8pt\normalsize1}}
\def\RR{{\rm I\kern-.2emR}}

\providecommand{\ignore}[1]{}

\newcommand{\scalarZ}{0}
\newcommand{\scalarO}{1}
\newcommand{\scalarPlus}{+}

\newcommand{\ip}[2]{\langle #1 ~|~ #2 \rangle}
\newcommand{\usat}{\texttt{UNIQUE-SAT}}
\newcommand{\boolt}{\textsf{Bool}} 

\newcommand{\btrue}{\texttt{\textbf{true}}}
\newcommand{\ff}[1]{\mathbb{F}_{#1}}
\newcommand{\todo}[1]{\textbf{TODO.~#1}}

\newcommand{\bitem}{\begin{itemize}}
\newcommand{\eitem}{\end{itemize}}
\newcommand{\benum}{\begin{enumerate}}
\newcommand{\eenum}{\end{enumerate}}
\newcommand{\beq}{\begin{equation}}
\newcommand{\eeq}{\end{equation}}
\newcommand{\beqa}{\begin{eqnarray}}
\newcommand{\eeqa}{\end{eqnarray}}
\newtheorem{definition}{Definition}

\newtheorem{proposition}{Proposition}

\newcommand{\bproof}{\begin{proof}}
\newcommand{\eproof}{\end{proof}}
\newcommand{\bprop}{\begin{proposition}}

\newcommand{\bdef}{\begin{definition}}

\input{Qcircuit}

\begin{document}
\title{The Power of Discrete Quantum Theories}
\author{Andrew J. Hanson, Gerardo Ortiz, Amr Sabry, and Jeremiah Willcock}
\affiliation{Quantum and Natural Computing Group, Indiana University, 
Bloomington IN 47405}
\begin{abstract}
%% 
%% The restrictions of quantum mechanics to finite fields produce stronger
%% (``supernatural'') theories. In the case of full quantum mechanics, the
%% parallel execution of superpositions can be done completely uniformly and
%% without any communication or synchronization between the various threads. But
%% in the finite field versions, it appears that one can single out some of the
%% threads and treat them specially which enables supernatural algorithms.
%% 
%We explore the implications of restricting the framework of quantum
%theory and quantum computation to finite fields.  The simplest proposed case appears to
%imply the intuitively unacceptable result that it is unnaturally
%strong, permitting the constant evaluation, polynomial time solution
%of SAT problems.  
%The next most general approach chooses finite fields with no solution
%to $x^2+1=0$, and thus permits an elegant complex representation of
%the extended field by adjoining $i=\sqrt{-1}$.  In these more natural
%finite-field quantum theories, it appears that one can single
%out some of database search cases and treat them specially, enabling
%``supernatural'' performance.  As the size of the field increases, the
%possibility of such phenomena decreases, and the properties of
%quantum theory with ordinary complex coefficients may be approached.
 We explore the implications of restricting the framework of quantum theory
 and quantum computation to finite fields.  The simplest proposed theory is
 defined over arbitrary finite fields and loses the notion of unitaries. This
 makes such theories unnaturally strong, permitting the search of
 unstructured databases faster than asymptotically possible in conventional
 quantum computing. The next most general approach chooses finite fields with
 no solution to $x^2+1=0$, and thus permits an elegant complex-like
 representation of the extended field by adjoining $i=\sqrt{-1}$. Quantum
 theories over these fields retain the notion of unitaries and --- for
 particular problem sizes --- allow the same algorithms as conventional
 quantum theory. These theories, however, still support unnaturally strong
 computations for certain problem sizes, but the possibility of such
 phenomena decreases as the size of the field increases.
\end{abstract}
\maketitle

%%%%%%%%%%%%%%%%%%%%%%%%%%%%%%%%%%%%%%%%%%%%%%%%%%%%%%%%%%%%%%%%%%%%%%%%

{\it Introduction.}  Quantum computing with complex coefficients technically
involves uncomputable numbers and unlimited resources.  Specifically, it is
well known that the set of computable complex numbers is countable whereas
the set of all complex numbers is uncountable.  Since we do not completely
understand the source of the extended power of conventional quantum
computation~\cite{qcdef}, it is therefore both interesting and potentially
important to investigate the possible origins of quantum computational
capacity.  Here we explore the remarkable properties that result when we
replace continuous complex numbers by appropriate finite fields.  This
apparently simple step adds new and bizarre properties to the well-known
post-classical computing power for which quantum computing has justifiably
attracted such attention.

We will show that, for finite fields of order $p^2$, with the prime $p$ of
the form $4 r + 3$ ($r$ a non-negative integer), the complex numbers have
extremely compelling and natural discrete analogs that permit essentially all
of the standard requirements of quantum computing to be preserved. Under
suitable conditions, we have amplitude-based partitions of unity, unitary
transformations, entanglement, and so forth.  What is new is that, because of
the cyclic nature of arithmetic in the finite complex field, excessive
computational power can result. We explore the mechanisms for these phenomena
and speculate on their implications. The circumstances in which such
supercomputation can occur depend on special numerical conditions that become
more and more scarce as the size of the finite field increases.  This leads
to the conjecture that, as the size of the field becomes large enough, most
of the properties of conventional quantum mechanics would be recovered.  This
leaves open the question of whether conventional quantum mechanics is
physical, or whether perhaps extremely large discrete quantum theories that
contain only computable numbers are at the heart of our physical universe.

%% What would happen if Nature discovers that most of the complex numbers are
%% uncomputable? There is a well established result in Mathematics that
%% indicates that the set of computable complex numbers is a set of zero measure
%% (?). If the number of resources we have is finite then it is clear that there
%% are limitations to ...  We all believe that there is a difference between
%% Nature and Mathematics and moreover, our experience indicates that our
%% abstract constructs based on infinite fields or continuum theories are more
%% than successful approximations to the description and control of our external
%% real world. So, why should we bother to even think about such a
%% question. Nature's messages through experimentation and Intellectual
%% curiosity are key to establish the laws of Nature but We would like, for
%% example, understand where the power of standard quantum computation is coming
%% from.
%%
%%In the present Letter we will answer the question: Can one systematically
%%approximate standard quantum theory by a discrete theory whose inner workings
%%are finite fields?  This is not merely an intellectual game. We will show
%%that if one is not careful enough one could build an in principle
%%``supernatural'' machine that would for instance find a pin in a haystack in
%%\emph{constant time} defeating the most powerful quantum computer. ...

%%%%%%%%%%%%%%%%%%%%%%%%%%%%%%%%%%%%%%%%%%%%%%%%%%%%%%%%%%%%%%%%%%%%%%%%
{\it Modal Quantum Theory.}  The traditional mathematical framework of
conventional quantum theory is that of Hilbert spaces over the field of
complex numbers.  Since this field is infinite, it is natural to ask whether
versions of quantum theory based on finite fields exist and can be used to
approximate conventional quantum theory, thus yielding insights into the
power of quantum computing.

Recently Schumacher and Westmoreland~\cite{modalqm} showed that it is
possible to define versions of quantum theory over finite fields, which they
call modal quantum theories. Such theories retain several key quantum
characteristics including notions of superposition, interference,
entanglement, and mixed states, along with time evolution using invertible
linear operators, complementarity of incompatible observables, exclusion of
local hidden variable theories, impossibility of cloning quantum states, and
the presence of natural counterparts of quantum information protocols such as
superdense coding and teleportation.  These modal theories are obtained by
collapsing the Hilbert space structure over the field of complex numbers to
that of a plain vector space over an \emph{arbitrary} finite field. In the
resulting structure, all non-zero vectors represent valid quantum states, and
the evolution of a closed quantum system is described by {\it any} invertible
linear map.

Specifically, consider a 1-qubit system with basis vectors $\ket{0}$
and~$\ket{1}$. In conventional quantum theory, there exist an infinite number
of states for a qubit of the form $\alpha\ket{0} + \beta\ket{1}$, with
$\alpha$ and $\beta$ elements of the underlying field of complex numbers
subject to the normalization condition $|\alpha|^2+|\beta|^2=1$. Moving to a
finite field immediately limits the set of possible states as the
coefficients~$\alpha$ and~$\beta$ are now drawn from a finite set. In
particular, in the field $\ff{2}=\{0,1\}$ of booleans, there are exactly four
possible vectors: the zero vector, the vector $\ket{0}$, the vector
$\ket{1}$, and the vector $\ket{0} + \ket{1}=\ket{+}$.  Since the zero vector
is considered non-physical, a 1-qubit system can be in one of only three
states.  The dynamic evolution of these 1-qubit states is described by any
invertible linear map, i.e., by any linear map that is guaranteed never to
produce the zero vector from a valid state. There are exactly 6 such maps:
\[\begin{array}{c}
X_0 = \begin{pmatrix}
1 & 0 \\
0 & 1 
\end{pmatrix} , 
\qquad\qquad
X_1 = \begin{pmatrix}
0 & 1 \\
1 & 0 
\end{pmatrix} , 
\\ \\
S = \begin{pmatrix}
1 & 0 \\
1 & 1 
\end{pmatrix} , 
\qquad
S^\dagger = \begin{pmatrix}
1 & 1 \\
0 & 1 
\end{pmatrix} , 
\qquad
\begin{pmatrix}
0 & 1 \\
1 & 1 
\end{pmatrix} , 
\  \
\begin{pmatrix}
1 & 1 \\
1 & 0 
\end{pmatrix} .
\end{array}\]
This space of maps is clearly quite impoverished compared to the full set of
1-qubit unitary maps in conventional quantum theory. In particular, it does
not include the Hadamard transformation. The space also includes non-unitary
maps such as $S$ and $S^\dagger$ that are not allowed in conventional quantum
computation.

Measurement in the standard basis is fairly straightforward: measuring
$\ket{0}$ or $\ket{1}$ deterministically produces the same state while
measuring $\ket{+}$ non-deterministically produces $\ket{0}$ or $\ket{1}$
with no particular probability distribution. In other bases, the measurement
process is complicated by the fact that the correspondence between
$\ket{\phi}$ and its {\it dual} $\bra{\phi}$ is basis-dependent and that the
underlying finite field is necessarily cyclic. For example, in the field of
booleans addition $(+)$ and multiplication $(*)$ are modulo 2, which means
that:
\begin{eqnarray}
\label{eqn}
\ip{+}{+} 
= 
(\scalarO * \scalarO) + (\scalarO * \scalarO) 
= 
\scalarO + \scalarO 
= 
\scalarZ .
\end{eqnarray}

%%%%%%%%%%%%%%%%%%%%%%%%%%%%%%%%%%%%%%%%%%%%%%%%%%%%%%%%%%%%%%%%%%%%%%%%
 {\it Modal Quantum Computing.}  Although modal quantum theories were
described as ``toy'' quantum theories, they appear to be endowed with
``supernatural'' powers.  We show next that it is possible --- in even the
simplest of modal theories --- to deterministically solve a black box version
of the \usat\ problem.  The \usat\ problem is that of deciding whether a
given boolean formula has a satisfying assignment, assuming that it has at
most one such assignment. Surprisingly, this problem is, in a precise
sense~\cite{Valiant198685}, just as hard as the general satisfiability
problem and hence all problems in the~NP complexity class. Our generalization
replaces the boolean formula with a arbitrary classical boolean function. A
solution to the generalized problem can be used to solve an unstructured
database search of size $N$ using $O(\log{N})$ black box evaluations by
binary search on the database. This algorithm outperforms the known
asymptotic bound $O(\sqrt{N})$ for unstructured database search in
conventional quantum computing.

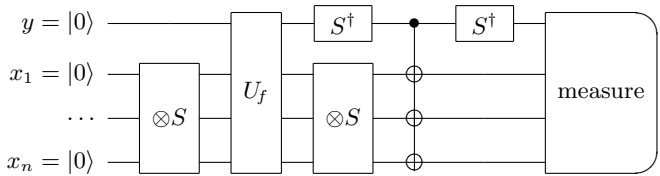
\begin{figure}[th]
\[\begin{array}{@{\!\!}c}
\hspace*{1.35cm} \Qcircuit @C=1.3em @R=.9em {
\lstick{y=\ket{0}}   & \qw                      & \multigate{3}{\uf} &  \gate{~S^\dagger~}         & \ctrl{3} & \gate{~S^\dagger~} & \multimeasureD{3}{\text{measure}} \\
\lstick{x_1=\ket{0}} & \multigate{2}{\otimes S} & \ghost{\uf}        &  \multigate{2}{\otimes S} & \targ    & \qw              & \ghost{\text{measure}}           \\
\lstick{\ldots}      & \ghost{\otimes S}        & \ghost{\uf}        &  \ghost{\otimes S}        & \targ    & \qw              & \ghost{\text{measure}}           \\
\lstick{x_n=\ket{0}} & \ghost{\otimes S}        & \ghost{\uf}        &  \ghost{\otimes S}        & \targ    & \qw              & 
\ghost{\text{measure}}           
}
\end{array}\]
\caption{Circuit for black box \usat\ in modal quantum theory over the field 
$\ff{2}$.  For notation see text.}
\label{fig:alg}
\end{figure}

Technically, consider a classical function $f : \boolt^n \rightarrow \boolt$
that takes $n$ bits and that returns at most one \btrue\ result. The
algorithm described below (and pictorially in Fig.~\ref{fig:alg}) takes as
input such a classical function and decides, deterministically and in a
constant number of black box evaluations, whether $f$ is satisfiable or not.

In the following, we use $\overline{x}$ to denote a sequence $x_1, x_2,
\ldots, x_n$ of $n$ bits. Given the function $f : \boolt^n \rightarrow
\boolt$, we construct the Deutsch quantum black box $\uf$ as follows
\cite{NCbook}: $ \uf \ket{y}\ket{\overline{x}} ~=~ \ket{y \scalarPlus
f(\overline{x})}\ket{\overline{x}} $.  The algorithm consists of the
following steps. (1) Initialize an $n+1$ qubit state to
$\ket{0}\ket{\overline{0}}$. (2) Apply the map~$S$ (defined in the previous
section) to each qubit in the second component of the state. (3) Apply the
quantum black box $\uf$ to the entire state. (4) Again apply the map $S$ to
each qubit in the second component of the state. (5) Apply the map
$S^\dagger$ to the first component of the state. (6) Conditional on the first
component of the state being $\ket{a}$, apply the map $X_a$ to each qubit in
the second component of the state where $X_0$ and $X_1$ are defined in the
previous section. (7) Again apply the map $S^\dagger$ to the first component
of the state. (8) Measure the resulting state in the standard basis for $n+1$
qubits. It is straightforward to calculate that if the measurement yields
$\ket{0}\ket{\overline{0}}$ then the function $f$ is \emph{unsatisfiable}. If
the measurement is anything else then the function $f$ is \emph{satisfiable}.

%%%%%%%%%%%%%%%%%%%%%%%%%%%%%%%%%%%%%%%%%%%%%%%%%%%%%%%%%%%%%%%%%%%%%%%%
{\it Discrete Quantum Theory. }  We propose variants of modal quantum
theories, which we call discrete quantum theories, that aim to exclude
``supernatural'' algorithms such as the one presented in the previous section
by retaining most of the structure of Hilbert spaces over the field of
complex numbers.
%% The structure which we aim to approximate consists of the following
%% components:
We wish to approximate as closely as possible the following features
of conventional quantum theory: (i) the
field of complex numbers, (ii) a vector space over the field of complex 
numbers, and (iii) an inner product $\ip{\psi}{\phi}$ associating a 
complex number to each pair of vectors that satisfies the following 
properties:
\begin{itemize}
\item[A.] $\ip{\phi}{\psi}$ is the complex conjugate of $\ip{\psi}{\phi}$; 
\item[B.] $\ip{\phi}{\psi}$ is conjugate linear in its first argument and
linear in its second argument;
\item[C.] $\ip{\phi}{\phi}$ is always non-negative and is equal to~0 only
if $\ket{\phi}$ is the zero vector. 
\end{itemize}

\begin{figure*}[htb]
 \begin{center}
\vspace{-0.2in}  
\includegraphics[width=2.0\columnwidth]{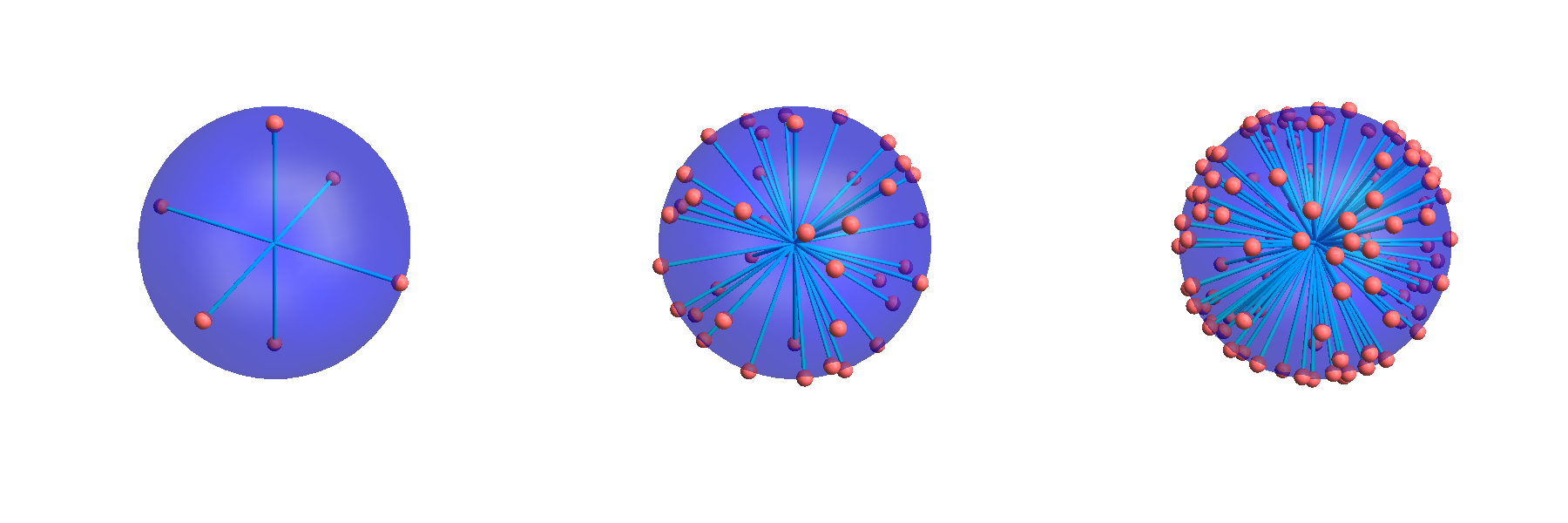}
\vspace{-0.7in}  %% adjust blank space?
 \end{center}
 \caption{The discrete versions of the  2-dimensional 1-qubit Hilbert
 space (the Bloch sphere) that are irreducible
distinct state vectors of unit norm in the field, with
the finite fields $\ff{p^2}$ for $p=3$, $p=7$, and $p=11$.
For example, in $\ff{3^2}$, there are 24 vectors of norm 1, but only 6
inequivalent vectors, as shown; the 4 equivalent vectors in each
class differ only by a discrete phase.}
 \label{fig_qm}
\end{figure*}

\paragraph*{Complex Numbers.}
Although arbitrary finite fields do not have enough structure to represent
approximations to the complex numbers, some finite fields do. To understand
this point, we review some known facts about finite fields. All finite fields
have sizes that are powers of primes. Thus in a field $\ff{q}$, the size $q$
must be of the form~$p^m$ where the prime number~$p$ is known as the
\emph{characteristic} of the field and where~$m$ is known as the
\emph{degree} of the field. The polynomial $x^2+1=0$ is \emph{irreducible}
over a prime field $\ff{p}$ with $p$ odd if and only if $p$ is of the
form~$4r+3$, with $r$ a non-negative integer.  In other words, the polynomial
is irreducible over $\ff{3}, \ff{7}, \ff{11}, \ff{19}, \ldots$~\cite{galois}.
Furthermore, any such field can be extended to a field~$\ff{p^2}$ whose
elements can be viewed as complex numbers with the real and imaginary parts
in~$\ff{p}$. In such a special field, the Frobenius automorphism of an
element (defined as $a^p$) agrees with the usual definition of complex
conjugation. Concretely, the field~$\ff{3^2}$ has 9 elements:
\[
0, 1, -1, i, 1+i, -1+i, -i, 1-i, -1-i \ ,
\]
which are all the complex numbers one can form using the integers modulo 3 as
real and imaginary coefficients. Similarly, the field $\ff{7^2}$ has 49
elements of the form $a+ib$ where $a,b$ are integers in the range $[-3,3]$
and addition and multiplication are modulo 7.

\paragraph*{Inner Products.} 
Consider a $d$-dimensional vector space over the fields $\ff{p^2}$ that
approximate the complex numbers where, in general, there is no connection
between the dimension of the vector space $d$ and the characteristic of the
field $p$. Let $\ket{\phi} = (a_0~a_1~\ldots~a_{d-1})^T$ and $\ket{\psi} = 
(b_0~b_1~\ldots~b_{d-1})^T$
%% \[
%% \ket{\phi} = \begin{pmatrix} a_0 \\ a_1 \\ \vdots \\ a_{d-1} \end{pmatrix} \qquad 
%% \ket{\psi} = \begin{pmatrix} b_0 \\ b_1 \\ \vdots \\ b_{d-1} \end{pmatrix} 
%% \]
with the scalars $a_j$ and $b_j$ drawn from the field elements and where
$(.)^T$ is the transpose. The Hermitian dot product of these vectors is:
\[
\ip{\phi}{\psi} = \sum_{j=0}^{d-1} ~a_j^p~b_j^{\;}
\]
This product satisfies conditions A and B for inner products. Condition C is
violated in every finite field as there always exists a non-zero vector
$\ket{\phi}$ such that $\ip{\phi}{\phi} = 0$. The reason is that addition in
finite fields eventually ``wraps around'' (because of their cyclic or modular
structure) making the notions of positive and negative meaningless and
allowing the sum of non-zero elements to be zero.  (See Eq.~(\ref{eqn}) at
the end of the first section.) This fact has non-trivial consequences.

\paragraph*{Postulates.}
Given that we can retain most of the structure of conventional quantum
theories in finite fields, the postulates of discrete quantum theory below
are almost identical to the usual ones. In more detail, the state space of an
isolated discrete quantum system is a vector space over a field that
approximates the complex numbers as shown above. In that space, complex
conjugates, unitaries, and Hermitian operators have the usual
definitions. Furthermore, (1) The state of an isolated system is described by
a vector $\ket{\psi}$ such that $\ip{\psi}{\psi}=1$, with vectors that differ
by a scalar multiplier identified; (2) The state space of a distinguishable
composite system is the tensor product of the component systems; (3)
Observable quantities are described by operators $O$ such that $O=O^\dagger$;
(4) The evolution of the system is described by unitary maps $U$; (5) If
$\ket{a}$ is an eigenvector of $O$ with eigenvalue $a$, i.e., if $O \ket{a}=a
\ket{a}$, then measurement of property $O$ realizes $a$. Because of the
cyclic nature of the field, traditional probability measures are not directly
applicable and some care is needed to define the probabilities of measurement
outcomes. We note that our algorithms for solving the black box \usat\ only
rely on probability measures distinguishing certain from impossible
events. 

Assuming the underlying field to be $\ff{3^2}$, there are exactly 6 1-qubit
state vectors, which appear as symmetric points on the Bloch sphere. We
observe that, for $\ff{p^2}$, there appear to be $p(p-1)$ unique unit norm
states on the discrete Bloch sphere, with $(p+1)$ equivalent discrete copies
(points on the circle realizing the discrete Hopf fibration) corresponding to
each unique state.  In Fig.~\ref{fig_qm} we plot these states on the Bloch
sphere for $p=3$, $7$, and~$11$.  Similar discrete maps can be performed for
$n$ qubit states.

%% \begin{figure*}[htb]
%%  \begin{center}
%% \includegraphics[width=2.0\columnwidth]{./picB.png}
%% \end{center}
%% \label{fig_qm2}
%% \end{figure*}

The unitary operators over the field $\ff{3^2}$ include the Hadamard
transformation: 
\[
H = (1+i) \begin{pmatrix} 1 & 1 \\ 1 & -1 \end{pmatrix}  ,
\]
whose rows and columns are mutually orthogonal unit vectors. Recall that, in
this field, $(1-i)(1+i) = 1 + 1 = -1$ and that $(-1) + (-1) = 1$ since all
operations are modulo 3. No-cloning, the fundamental tenet underlying
conventional quantum theory, is respected in our discrete quantum theory.

Any observable $O$ in a 1-qubit space can be written as:
\[
O=\sum_{\mu=0}^3 a_\mu \, X_\mu =
\begin{pmatrix}
a_0 + a_3 & a_1 - i a_2 \\
a_1 + i a_2 & a_0 - a_3 
\end{pmatrix}
\]
where $X_0$, $X_1$ were defined above, and the remaining Pauli matrices 
$X_2,X_3$ allowed in the field are: 
\[\begin{array}{c}
X_2 = \begin{pmatrix}
0 & -i \\
i & 0 
\end{pmatrix} , 
\qquad\qquad
X_3 = \begin{pmatrix}
1 & 0 \\
0 & -1 
\end{pmatrix} . 
\end{array}
\]
Since $O=O^\dagger$ this implies that its diagonal elements can only be
$0,1,-1$, while the off-diagonal ones can be any of the 9 field elements.
Incompatibility, i.e., non-commutativity, of observables leads to Heisenberg
uncertainty relations among them, a consequence of the validity of Schwarz's
inequality in discrete quantum theory.

%% \todo{Delete this comment:} I checked and these are the four observables:
%% \[\begin{array}{c}
%% \begin{pmatrix}
%% 1 & 0 \\
%% 0 & 1 
%% \end{pmatrix} , 
%% \quad
%% \begin{pmatrix}
%% 0 & 1 \\
%% 1 & 0 
%% \end{pmatrix} , 
%% \quad
%% \begin{pmatrix}
%% 0 & i \\
%% -i & 0 
%% \end{pmatrix} , 
%% \quad
%% \begin{pmatrix}
%% 1 & 0 \\
%% 0 & -1 
%% \end{pmatrix} 
%% \end{array}\]

%%%%%%%%%%%%%%%%%%%%%%%%%%%%%%%%%%%%%%%%%%%%%%%%%%%%%%%%%%%%%%%%%%%%%%%%
{\it Discrete Quantum Computing. }  As shown above, by choosing particular
finite fields, it is possible to retain all the structure of conventional
quantum theory except for condition~C of inner products. The smallest
field~$\ff{3^2}$ already has enough structure to express the standard
Deutsch-Jozsa~\cite{NCbook} algorithm as this algorithm only requires
normalized versions of vectors or matrices with the scalars $0$, $1$, and
$-1$. More complex algorithms such as Grover's database search or Shor's
period finding~\cite{NCbook} also work as expected in ``large enough'' finite
fields. Consider the diffusion matrix for searching an unstructured database
of size~$N$ using Grover's algorithm:
\[
\begin{pmatrix}
-1 + 2/N & 2/N & 2/N & \ldots & 2/N \\
2/N & -1 + 2/N & 2/N & \ldots & 2/N \\
2/N & 2/N & -1 + 2/N & \ldots & 2/N \\
\ldots & \ldots & \ldots & \ldots & \ldots \\
2/N & 2/N & 2/N & \ldots & -1 + 2/N
\end{pmatrix}
\]
This diffusion matrix is applied $\sqrt{N}$ times. This means that if $N=4$,
the algorithm needs to be expressed using probability amplitudes of the form
$\pm\frac{1}{2}$. In the field~$\ff{3^2}$, these amplitudes collapse to $\pm
1$ and the algorithm fails to work. However in the field $\ff{7^2}$, these
amplitudes can properly be expressed and the algorithm works as
expected. Generally as the size of the database grows, the size of the
underlying field must grow proportionally.

%% Indeed, discrete quantum computing is restricted to unitary gates
%% as opposed to modal quantum computing.  The interesting situation for when
%% the finite field is too small to represent the structure of the problem is
%% discussed in the next section.
%% 
%% \todo{The above should be checked and worked out in full detail detail. It
%% would be good to have a precise mathematical formula relating the size of the
%% database to the size of the field. Perhaps we should discuss other algorithms
%% including Shor's.}

%%%%%%%%%%%%%%%%%%%%%%%%%%%%%%%%%%%%%%%%%%%%%%%%%%%%%%%%%%%%%%%%%%%%%%%%
\begin{figure}[t]
\[\begin{array}{c}
\hspace*{1.5cm} \Qcircuit @C=1.7em @R=.9em {
\lstick{y=\ket{0}}   & \qw              & \multigate{3}{\uf} &  \qw              & \multimeasureD{3}{\text{measure}} \\
\lstick{x_1=\ket{0}} & \multigate{2}{H} & \ghost{\uf}        &  \multigate{2}{H} & \ghost{\text{measure}}           \\
\lstick{\ldots}      & \ghost{H}        & \ghost{\uf}        &  \ghost{H}        & \ghost{\text{measure}}           \\
\lstick{x_N=\ket{0}} & \ghost{H}        & \ghost{\uf}        &  \ghost{H}        & \ghost{\text{measure}}           
}
\end{array}\]
\caption{Circuit for black box \usat\ in discrete quantum theories.}
\label{fig:dbsearch}
\end{figure}
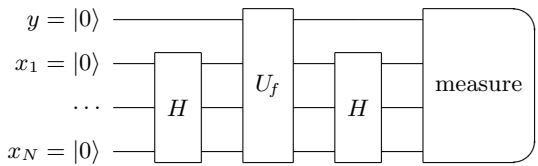

{\it Building a Discrete Quantum Computer.} We have argued above that if the
field is large enough, then discrete quantum computing approaches
conventional quantum computing and that if the size of the field is too
small, then the usual algorithms fail to work. However it is possible, in
some situations, to exploit the cyclic behavior of the field to creatively
cancel probability amplitudes and solve problems with what again appears to
be ``supernatural'' efficiency. We illustrate this behavior with the
algorithm in Fig.~\ref{fig:dbsearch}, which is a variant of the algorithm in
Fig.~\ref{fig:alg}. However, unlike the situation in modal quantum theories,
the algorithm does not always succeed deterministically using a constant
number of black box evaluations. This supernatural behavior only happens if
the characteristic $p$ of the field divides $2^N - 1$. For a database of
fixed size, this match becomes less likely as the size of the field
increases. For a given field, it is possible to expand any database with
dummy records to satisfy the divisibility property. 

\paragraph*{Physical connections.} 
We conclude by pointing out a connection between our discrete quantum
theories and Schwinger's foundational attempts to formulate quantum mechanics
from measurement \cite{Sbook,vourdas}.  He proposed a measurement algebra
derived from a selected set of experiments, including coordinates and momenta
whose eigenvalues are modular integers.  Interestingly, although this
formulation shares with the discrete quantum theories the cyclic structure
induced by the finite fields, it differs in that the infinite complex numbers
fields are used to define a state space that is a Hilbert space with a
standard inner product.  Related situations also appear in quantum field
theories,
%% such as quantum electrodynamics %%
where competition between confinement and deconfinement of elementary
particles may appear depending upon the compact ({\it cyclic\/}) or
non-compact nature of the quantum fields.

%%%%%%%%%%%%%%%%%%%%%%%%%%%%%%%%%%%%%%%%%%%%%%%%%%%%%%%%%%%%%%%%%%%%%%%%
{\it Acknowledgments.}  We would like to thank J. R. Busemeyer,
J. M. Dunn, A.  Lumsdaine, and L. S. Moss for many inspiring discussions. We
acknowledge support from Indiana University's Institute for Advanced Study.

\vspace*{-0.25in}

%%%%%%%%%%%%%%%%%%%%%%%%%%%%%%%%%%%%%%%%%%%%%%%%%%%%%%%%%%%%%%%%%%%%%%%%
%% \bibliography{p.bib}

\begin{thebibliography}{99}
\bibitem{qcdef} We use the phrase ``conventional quantum theory'' where
  necessary to distinguish the usual quantum theory and quantum computing
  paradigm using (continuous) complex numbers from discrete quantum theory.
  Alternative terminology in the literature includes ``actual,''
  ``standard,'' and ``ordinary'' quantum theory.
\bibitem{galois}
Fields $\ff{q}$ where $q$ is a power of a prime $p$, i.e.,
$q=p^m$, are  known as Galois fields.
\end{thebibliography}
%% \end{document}

\end{document}